\documentclass[useAMS,usegraphicx]{mn2e}
\usepackage{rotate}
\usepackage{times}
\newif\ifAMStwofonts
\AMStwofontstrue

\def\gs{\mathrel{\hbox{\rlap{\hbox{\lower4pt\hbox{$\sim$}}}\hbox{$>$}}}}
\def\ls{\mathrel{\hbox{\rlap{\hbox{\lower4pt\hbox{$\sim$}}}\hbox{$<$}}}}

\def\xmm{{\it XMM-Newton}}

\def\et{{et al.\ }}
\def\mcg{{MCG--6-30-15}}
\def\nab{{NAB~0205+024}}
\def\3c{{3C~273}}

\def\nh{{N_{\rm H}}}

\def\arcs{{\hbox{$^{\prime\prime}$}}}

\def\A{{\rm\thinspace \AA}}
\def\cm{{\rm\thinspace cm}}
\def\erg{{\rm\thinspace erg}}

\def\ga{{\rm\thinspace gauss}}

\def\keV{{\rm\thinspace keV}}

\def\s{{\rm\thinspace s}}
\def\ks{{\rm\thinspace ks}}

\def\ergpscmps{\hbox{$\erg\cm^{-2}\s^{-1}\,$}}
\def\ergpscmpspa{\hbox{$\erg\s^{-1}\cm^{-2}\A^{-1}\,$}}
\def\ergps{\hbox{$\erg\s^{-1}\,$}}

\def\pscm{\hbox{$\cm^{-2}\,$}}

\title[\nab\ observed with \xmm]
      {
The narrow-line quasar \nab\ observed with \xmm
      }

\author[L. C. Gallo et al.]
       {L. C. Gallo,$^1$
	Th. Boller,$^1$
	W. N. Brandt,$^2$
	A. C. Fabian,$^3$
	S. Vaughan$^{4}$  \\
$^{1}$ Max-Planck-Institut f\"ur extraterrestrische Physik, Postfach 1312, 85741 Garching, Germany \\
$^{2}$ Department of Astronomy and Astrophysics, The Pennsylvania State University, 525 Davey Lab, University Park, PA 16802, USA\\
$^{3}$ Institute of Astronomy, University of Cambridge, Madingley Road, Cambridge CB3 0HA\\
$^{4}$ X-Ray and Observational Astronomy Group, Department of Physics and Astronomy, University of Leicester, Leicester LE1 7RH
}
\date{Accepted. Received. }
\pagerange{\pageref{firstpage}--\pageref{lastpage}}
\pubyear{2004}
\begin{document}
\maketitle
\label{firstpage}

\begin{abstract}
The \xmm\ observation of the narrow-line quasar \nab\ reveals three
striking differences since it was last observed in the X-rays with
$ASCA$.  Firstly, the 2--10~keV power-law is notably steeper.
Secondly, a hard X-ray flare is detected, very similar to that seen in 
I~Zw~1. Thirdly, a strong and broad emission feature is detected   
with the bulk of its emission redward of 6.4~keV, and extending
down to $\sim 5$~keV in the rest frame.
The most likely explanation for the broad feature is neutral iron emission
emitted from a narrow annulus of an accretion disc close to the black hole.
The hard X-ray flare could be the mechanism which illuminates this region
of the disc, allowing for the emission line to be detected.
The combination of effects can be understood in terms of 
the `thundercloud' model proposed by Merloni \& Fabian.
 
\end{abstract}


\section{Introduction}
\label{sect:intro}
Narrow-line Seyfert 1 galaxies (NLS1) offer an extreme view of the
Seyfert~1 phenomenon.  For example, these objects show:
\begin{itemize}

\item
The narrowest permitted optical lines, strongest optical 
Fe~\textsc{ii} emission,
and weakest [O~\textsc{iii}] emission (e.g. Osterbrock \& Pogge 1985;
Boroson \& Green 1992; Grupe 2004).

\item
Strong soft X-ray excesses (e.g. Boller, Brandt \&
Fink 1996; Leighly 1999a; Vaughan \et 1999) above a relatively steep hard
power-law (e.g. Brandt, Mathur \& Elvis 1999); 

\item
Large-amplitude X-ray variations on hourly time scales 
(e.g. Boller, Brandt \& Fink 1996; Leighly 1999b).

\item
Weak or negligible optical variability (e.g. Klimek, Gaskell \& Hedrick
2004).

\end{itemize}
Many of the properties can be explained in terms of a high accretion rate
and relatively low-mass black hole (e.g. Sulentic \et 2000; Boroson 2002;
Grupe 2004).
These characteristics are not exclusive to lower-luminosity AGN
(i.e. Seyfert~1s), but are also seen in some quasar-type objects.
The most famous of these {\it narrow-line quasars} is the prototype
I~Zw~1, but the class also includes members such as PHL~1092, E~1346+266,
IRAS~13349+2438, and the {\em Neta A. Bahcall} object \nab~(Bahcall, Bahcall
\& Schmidt 1973).

\nab\ (Mrk~586; $z = 0.155$) has been the focus of previous X-ray studies with
$ASCA$ (e.g. Fiore \et 1998), and $ROSAT$ (Fiore \et 1994).  The strong soft 
excess and rapid variability (by factors of two on time scales of hours) have
made it an important object to study the nature of 
NLS1 behaviour in luminous narrow-line quasars.

In this paper we discuss the X-ray analysis of 
\nab\ as observed with \xmm.
The X-ray brightness of \nab\ (compared to other similar objects) 
affords us
the opportunity to study the NLS1 phenomenon in a quasar with relatively
high signal-to-noise.


\section{Observations and data reduction}
\label{sect:data}
\nab\ was observed with {\em XMM-Newton} (Jansen et al. 2001) for 
$50$~ks on 2002 July 23 (revolution 0480).
During this time the EPIC pn (Str\"uder et al. 2001)
and MOS (MOS1 and MOS2; Turner et al. 2001) cameras, as well
as the Optical Monitor (OM; Mason et al. 2001) and the Reflection Grating
Spectrometers (RGS1 and RGS2; den Herder et al. 2001) collected data.
The EPIC pn and MOS2 cameras were operated in full-frame mode and 
utilised the thin filter.  
The MOS1 camera was operated in timing mode, and the data will not be 
included in the current analysis.

The Observation Data Files (ODFs)
were processed to produce calibrated event lists using the {\em
XMM-Newton} Science Analysis System ({\tt SAS v6.0.0}). Unwanted hot,
dead, or flickering pixels were removed as were events due to
electronic noise.  Event energies were corrected for charge-transfer
losses, and EPIC response matrices were generated using the {\tt SAS} tasks
{\tt ARFGEN} and {\tt RMFGEN}.
Light curves were extracted from these event lists to
search for periods of high background flaring.
High-energy background flaring was substantial.
The total good exposure times selected for the pn and MOS2
were 20388~s and 31545~s, respectively.
The source plus background photons were extracted from a
circular region with a radius of 35$^{\prime\prime}$, and the background was
selected from an off-source region with a radius of 50$^{\prime\prime}$ 
and appropriately scaled to the source
region.  Single and double events were selected for the pn
detector, and single-quadruple events were selected for the MOS2.
Pile-up effects were determined to be negligible in the time-averaged data.
However, there was mild pile-up during a $\sim 4\ks$ period when the
source flux increased by $50\%$ (Sect. 4).  As a consistency check
the inner $5\arcs$ of the source extraction region was removed, thus correcting
the pile-up effect during the flare.  The step proved to be unnecessary, as
it had little effect on the spectral variability results (Sect. 4.2).
The total numbers of source plus background counts collected
in the 0.3--10~keV range by the pn and MOS2 instruments were 68372 and
29354, respectively.  The total numbers of counts collected from
the scaled background region were 1187 for the pn, and 1132 for the MOS2.
The {\em XMM-Newton} observation provides a substantial improvement in spectral
quality over the 51.8~ks $ASCA$-SIS exposure (118.6~ks duration) in
which $\approx 18130$ counts were collected (Leighly 1999b).  

The RGS were operated in standard Spectro+Q mode.
The first-order RGS spectra were extracted using the {\tt SAS} task 
{\tt RGSPROC}, and the response matrices were generated using 
{\tt RGSRMFGEN}. 
The total exposure times utilised for the analysis
were 46078~s and 44717~s for the RGS1 and RGS2, respectively.
The total number of source counts in the 0.35--1.5~keV range were
approximately 4915 (RGS1) and 4708 (RGS2).

The OM was operated in imaging mode for the entire observation.
In total, thirty-one 1000~s images were taken in three filters:
21 in $U$ ($3000-3900 \A$), and 5 in both $B$ ($3900-4900 \A$) and
$UVW2$ ($1800-2250 \A$).  The average apparent magnitude in each filter was
$U = 14.614 \pm 0.002$, $B = 15.701 \pm 0.004$, and $UVW2 = 14.126 \pm
0.011$.


\section{Spectral analysis}
\label{sect:fit}
Each of the EPIC spectra was compared to the
respective background spectrum to determine the energy range in which the 
source was reasonably detected above the background.
The source was detected above the background up to an observed energy of 
$\sim 10$~keV,
and $\sim$ 7~keV in the pn and MOS2 data, respectively.
The MOS data were ignored below 0.5~keV due to the
uncertainties
in the low-energy redistribution characteristics of the cameras (Kirsch
2003).
Therefore, the pn data between 0.3--10~keV and the MOS2 data between
0.5--7~keV were utilised during the spectral fitting, but the residuals from 
each instrument were examined separately to judge any inconsistency.
In addition the RGS data between 0.35--1.5~keV were also examined.

The source spectra were grouped such that each bin contained at least 20
counts. Spectral fitting was performed using {\tt XSPEC v11.3.0} (Arnaud
1996).
Fit parameters are reported in the rest-frame of the object, although the
figures remain in the observed-frame.
The quoted errors on the model parameters correspond to a 90\% confidence
level for one interesting parameter (i.e. a $\Delta\chi^2$ = 2.7 criterion).
Luminosities were derived assuming isotropic emission.
The Galactic column density toward \nab\ is 
$\nh = 3.51 \times 10^{20}\pscm$ (Dickey \& Lockman 1990).
A value for the Hubble constant of $H_0$=$\rm 70\ km\ s^{-1}\ Mpc^{-1}$ and
a standard cosmology with $\Omega_{M}$ = 0.3 and $\Omega_\Lambda$ = 0.7
was adopted.


\subsection{The broad-band X-ray continuum}

\label{sect:bbc}

\begin{figure}
\rotatebox{270}
{\scalebox{0.32}{\includegraphics{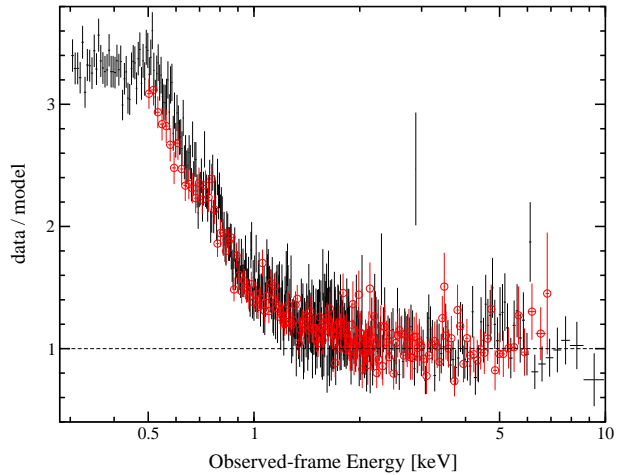}}}
\caption{The ratio (data/model) resulting from fitting an absorbed power-law
($\Gamma_{pn} = 2.26 \pm 0.07$ and $\Gamma_{MOS2} = 2.20 \pm 0.10$) to the
2--10~keV EPIC data and extrapolating to
lower energies.
The black crosses and red open circles are the pn and MOS2 residuals, 
respectively.  The apparently ``wild'' single data point at $E \approx 3$~keV
is less than $3\sigma$ above the ratio, and probably is just a statistical 
fluctuation.
The data have been binned for display purposes.
}
\label{fig:po}
\end{figure}

A single power-law with Galactic absorption was a poor fit to the EPIC data 
between 0.3--10~keV
($\chi^2 = 1937.7/738~dof$).  The higher statistics in the low-energy range
dominate the fit resulting in large residuals at higher energies which
demonstrates the need for multiple continuum components.
Fitting only the EPIC data above 2~keV with an absorbed power-law resulted in a
good fit ($\chi^2 = 272.4/302~dof$) and demonstrated agreement in the
photon indices measured by each instrument ($\Gamma_{pn} = 2.26 \pm 0.07$ and
$\Gamma_{MOS2} = 2.20 \pm 0.10$).  Extrapolating this model to lower energies
revealed a strong soft excess above the power-law continuum 
(Figure~\ref{fig:po}).

To investigate the nature of the continuum emission additional components
were included in the basic power-law fit to model the soft excess.
The data from each instrument were modelled separately in order to compare
the results.  In the case of the RGS, the high-energy model component (i.e.
the power-law) was kept fixed to the pn values.

\begin{figure}
\rotatebox{270}
{\scalebox{0.32}{\includegraphics{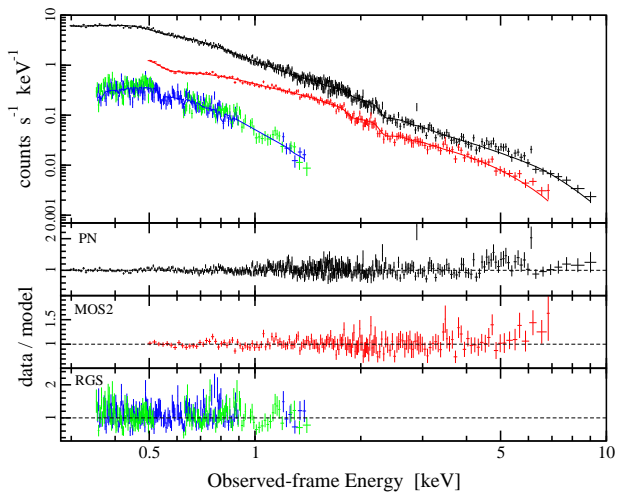}}}
\caption{The best-fit continuum model (blackbody plus power-law) fitted
to the \xmm\ data.  In the top panel the fit and data are shown for the
pn (upper curve; black), MOS2 (middle curve; red), and RGS (lower curve; blue
and green).  In the remaining panels the fit residuals (data/model) are shown 
for each instrument.
}
\label{fig:cont}
\end{figure}

To fit the broad-band continuum a number of models were used including
(i) a blackbody plus power-law, (ii) a double power-law and, (iii) a 
broken power-law.
The addition of a second continuum component was a substantial improvement
to the initial power-law fit; the best fit was obtained with model (i)
(Figure~\ref{fig:cont}).  In this case the 
blackbody temperature was $kT \approx 120$~eV, and the intrinsic absorption 
was negligible.  The 0.3--10\keV~flux and luminosity, corrected for
Galactic absorption, were $8.6 \times 10^{-12} \ergpscmps$ and 
$6.2 \times 10^{44} \ergps$, respectively.  The 2--10\keV~luminosity
was $1.1 \times 10^{44} \ergps$.
In Table~\ref{tab:cont} the model parameters for each fit are given.

\begin{table*}
\begin{center}
\caption{X-ray continuum models. The data from each EPIC instrument were modelled separately, but the
RGS1 and RGS2 data were fitted together.
The superscript $f$ marks the RGS parameters which are fixed to the best-fit
pn values.
The three continuum models, shown in column (1), are: 
(i) blackbody plus power-law, (ii) double power-law, and (iii) broken power-law.
In column (2) we indicate the instrument used to collect the data, and in 
column (3) the fit quality ($\chi^2_\nu$) is given.  Column (4) gives the measured intrinsic
cold absorption.  Column (5) is the temperature of the blackbody component.
Columns (6) and (7) are the photon indices of the power-law components (two
power-laws are required for models (ii) and (iii)).  Column (8) is the energy at
which the photon index changes between $\Gamma_1$ and $\Gamma_2$ in model (iii).
All models have been modified by line-of-sight Galactic absorption
(3.51 $\times$ 10$^{20}$ cm$^{-2}$).
}
\begin{tabular}{cccccccccc}                
\hline
(1) & (2) & (3) & (4)  & (5) & (6) & (7) & (8)  \\
 Continuum   &  Instrument  &  $\chi^2_\nu$    &   $\nh$           &   $kT$    &    $\Gamma_1$    &    $\Gamma_2$    & $E_{brk}$  \\
 Model       & 	            &   (dof)          & (10$^{20} \pscm$) &   (eV)    &                  &            & (keV) \\
\hline
 (i)         &  pn   	 & 1.04 (523) & $< 0.3$  & $121 \pm 2$ & $2.41 \pm 0.04$ & &  \\
	     &  MOS2  & 0.88 (209) &  $< 1.4$ & $118^{+5}_{-6}$ & $2.37^{+0.06}_{-0.05}$   \\
	     &  RGS  & 1.19 (313)  & $< 0.6$ & $127 \pm 4$ &  $2.41^{f}$  \\

 (ii)        &  pn & 1.76 (523) & $5.85^{+0.08}_{-0.06}$ & & $3.81 \pm 0.16$ & $1.54 \pm 0.13$ &     \\
	     &  MOS2 & 0.88 (209) & $15.9^{+5}_{-13}$ & & $5.74^{+0.52}_{-1.45}$ & $2.25^{+0.06}_{-0.21}$ &       \\
	     &  RGS  & 1.41 (313) & $< 1.9$ & & $3.16^{+17}_{-7}$ & $1.54^{f}$ &      \\

 (iii)       &  pn   & 1.54 (523) & $6.65^{+1.44}_{-1.18}$ &  & $3.74^{+0.12}_{-0.10}$  & $2.42 \pm 0.05$ & $1.38 \pm 0.04$    \\
	     &  MOS2 & 0.92 (209) & $< 2.5$ & & $3.34 \pm 0.06$ & $2.39 \pm 0.06$ & $1.22 \pm 0.08$      \\
	     &  RGS  & 1.36 (313) & $< 2.1$ & & $3.05^{+17}_{-7}$ & $2.42^{f}$ & $1.38^{f}$      \\
\hline
\label{tab:cont}
\end{tabular}
\end{center}
\end{table*}


\subsection{The RGS data}
\label{sect:rgs}

The RGS data were examined with finer energy bins to search for narrow
absorption and emission features which may be expected from a warm medium in 
an AGN.  The blackbody plus power-law continuum (model (i) in 
Table~\ref{tab:cont}) was adopted.  Within the uncertainties no strong narrow
features were revealed.  The most significant feature was a dip below the 
continuum at about 1~keV (see the lower panel of Figure~\ref{fig:cont}).
Adding a Gaussian absorption profile to model this feature
was an improvement to the RGS continuum model ($\Delta\chi^2 = 10.4$ for 3
additional free parameters).  The line had an intrinsic energy, width, and
equivalent width of $E = 1.19 \pm 0.03$~keV, $\sigma = 30^{+30}_{-15}$~eV,
$EW = -22.3^{+0.7}_{-0.9}$~eV, respectively.  
The measured energy is consistent with
Fe~L resonant absorption (e.g. Nicastro, Fiore \& Matt 1999), but the 
strength of the feature is about five times weaker than predicted for 
an IRAS~13224--3809 type absorber (see Nicastro \et for details).
The reliability of the detection is questionable given that
the absorption feature was
not detected in either of the EPIC spectra.  In addition, only the RGS2 is
operational between $\sim 0.9-1.2$~keV, due to a malfunctioning CCD in the
RGS1 at this energy range.  The feature could also be a calibration effect
as the flux calibration in the RGS is known to be uncertain by about 5\%
across the RGS band (Pollock 2003).


\subsection{A high-energy feature}
\label{sect:hef}

Clearly visible in the pn spectrum at $\sim 5$~keV 
(top two panels of Figure~\ref{fig:cont}) is an excess in the residuals which
could indicate the presence of an emission line.  
Indeed the addition of a
Gaussian profile was a significant improvement to the pn 0.3--10~keV fit (i)
from Table~\ref{tab:cont} ($\Delta\chi^2 = 20.3$ for 3 additional free 
parameters); however, the best-fit line energy of $E = 5.87^{+0.29}_{-0.27}$~keV
was inconsistent with iron emission, and no other elements in this spectral 
range are expected to produce such a wide and strong line 
($\sigma = 623^{+300}_{-237}$~eV; $EW = 515 \pm 6$~eV).
\begin{figure}
\rotatebox{270}
{\scalebox{0.32}{\includegraphics{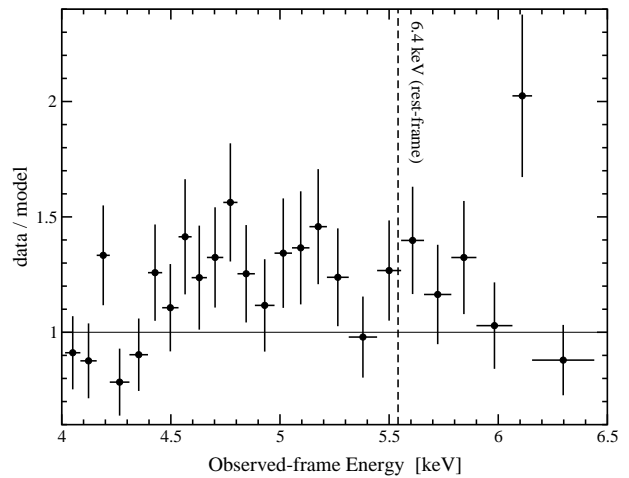}}}
\caption{The pn residuals (data/model) in the intrinsic
4.6--7.5~keV range (a blow-up of the second top panel in Figure~\ref{fig:cont}).
The dashed vertical line is the 6.4~keV rest energy.
}
\label{fig:line}
\end{figure}

Since the excess is strongly concentrated to the red side of 6.4~keV
(Figure~\ref{fig:line}), relativistic effects were considered.
The possibility that the emission feature arises from a disc 
around a Schwarzschild black hole (Fabian et al. 1989) was examined.  
The fit using the disc line was comparable to the Gaussian profile 
($\Delta\chi^2 = 20.3$ for 3 additional 
free parameters).  The best-fit line energy was $E = 5.69^{+0.15}_{-0.10}$~keV 
with an equivalent width of $EW = 455 \pm 5$~eV. 
The shift of the line to energies lower than $6.4$~keV suggests that 
gravitational redshift effects may be dominating the emission.
Given that the best-fit line energy did not correspond to iron emission,
we fixed the energy to 6.4~keV and refit the disc line model by allowing
the inner and outer disc radii to be free parameters.  This fit was
also acceptable ($\chi^2 = 525.8$ for $519~dof$), and indicated that
the disc line was emitted from a small annulus between about 18--24~$R_g$.
An annular emission region could be realised if a specific region of the disc 
is being
illuminated for some period of time (e.g. Iwasawa \et 1999).
Iron emission from an accretion disc around a Kerr black hole (Laor 1991) 
was also acceptable 
($\Delta\chi^2 = 16.1$ for 3 additional free parameters).
The line energy and inner radius were fixed at $E = 6.4$~keV
and $R_{in} = 1.235~R_g$ 
(the innermost stable orbit around a Kerr black 
hole).  
In this case only a lower-limit for the outer radius
was measured ($R_{out} > 14~R_g$). 
The best-fit line parameters are given in Table~\ref{table:line}.

Recent observations of some Seyferts have revealed narrow emission lines 
at peculiar energies redward of 6.4~keV (Turner et al. 2004; 
Yaqoob et al. 2003; Guainazzi 2003; Turner et al. 2002).
Dov\v ciak et al. (2004) explain these features as iron lines produced 
in a small range of radii by localised flares.
The feature in \nab\ could also be fitted with a series of three 
emission lines with energies of $E \approx 5.45, 5.95, 6.59$~keV and
unconstrained widths 
($\Delta\chi^2 = 28.6$ for 9 additional free parameters).
The lines were much stronger ($EW \approx 97-240$~eV) than the narrow lines 
discussed by the above
authors which have typical equivalent widths of a few tens of eV. 
In addition, splitting the broad feature into three components was not
required on a statistical basis.

A broad Gaussian emission profile was not statistically required by the
MOS2 data; however the MOS2 data were consistent with the pn model.
The broad feature was not detected by $ASCA$.  Simulations of the $ASCA$
observation using the pn model show that this was a signal-to-noise issue. 
 
\begin{table}
\caption{The best-fit line model parameters to the broad-band pn spectrum.
For each model the parameters given (as required) are: the power-law
photon index ($\Gamma$); line energy ($E$); line width ($\sigma$);
disc inclination ($i$); inner ($R_{in}$) and outer ($R_{out}$) disc radius 
in units of gravitational radii ($R_{g} = Gm/c^{2}$); power-law dependency
on the disc emissivity ($q$);
equivalent width ($EW$); flux ($F$) in units of $\times 10^{-14} \ergpscmps$.
Parameters marked with the superscript $f$ are fixed as leaving them free
did not improve the fits.
}
\centering
\label{table:line}            
\begin{tabular}{cl}
\hline
Gaussian & $\chi^2_\nu = 1.01 (520)$\\
 & $\Gamma = 2.47 \pm 0.02$ \\
 & $E = 5.87^{+0.29}_{-0.27}$~keV \\
 & $\sigma = 623^{+300}_{-237}$~eV \\
 & $EW = 515 \pm 6$~eV; $F = 8.46$ \\
\hline
Disc-line & $\chi^2_\nu = 1.01 (519)$ \\
 & $\Gamma = 2.45 \pm 0.03$ \\
 & $E^{f} = 6.4$~keV \\
 & $q^{f} = -2$; $R_{in} = 18^{+7}_{-12}~R_{g}$; $R_{out} = 24^{+17}_{-6}~R_{g}$ \\ 
 & $i = 31^{+5}_{-11}~degr$ \\
 & $EW = 490 \pm 11$~eV; $F = 6.10$ \\
\hline
Laor line & $\chi^2_\nu = 1.01 (520)$ \\
 & $\Gamma = 2.47 \pm 0.02$ \\
 & $E^{f} = 6.4$~keV \\
 & $q^{f} = 3$; $R^{f}_{in} = 1.235~R_{g}$; $R_{out} > 14~R_{g}$ \\
 & $i = 37^{+9}_{-14}~degr$ \\
 & $EW = 868 \pm 10$~eV; $F = 9.13$ \\
\hline
\end{tabular}
\end{table}


\section{Timing Analysis}
\label{sect:time}

\subsection{X-ray and optical light curves}
\label{sect:xolc}

\begin{figure}
\rotatebox{270}
{\scalebox{0.32}{\includegraphics{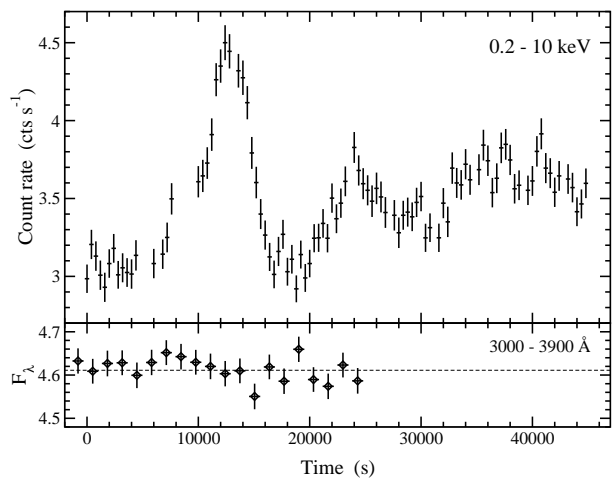}}}
\caption{
Top panel: The 0.2--10~keV light curve over the duration of the observation
using 400~s bins. The time axis is defined from the start of the pn 
observation at 07:37:55 on 2002-07-23 UT.
Bottom panel:  The $U$-filter 
light curve obtained with the OM during the first $\sim 25$~ks of the 
observation.  
Flux-density units are $\times 10^{-15} \ergpscmpspa$.
A dashed line is drawn at the average flux density.
The time axis is normalised to the pn start time.
}
\label{fig:lc}
\end{figure}

The 0.2--10~keV light curve of \nab\ is displayed in the top panel of 
Figure~\ref{fig:lc} (with $1\sigma$ error bars).  
Since the broad-band light curve is less sensitive to calibration
uncertainties the data between 0.2--0.3~keV have been included.
In addition, the light curve from the entire observation is shown since
the source is sufficiently brighter that the background (by about a factor
of ten), even during the background flaring.
The average count rate is $3.49 \pm 0.10$~counts s$^{-1}$.
The most striking feature is a flaring event about
10~ks into the observation in which the broad-band count rate increases by 
$\sim 50$\% in a few thousand seconds.
Averaging over three data points during flare maximum at 
$\sim 12.4$~ks and the three lowest data points when the flare achieves minimum
intensity at $\sim 16.8$~ks results in a count-rate change of 1.33 counts 
s$^{-1}$ in 3800~s (rest-frame).
Model (i) from Table~\ref{tab:cont} is extrapolated down to 0.2~keV to estimate
a 0.2--10~keV intrinsic luminosity of $7.74 \times 10^{44} \ergps$.
Thus the calculated change in count rate corresponds to a luminosity 
change of $\Delta$L$ = 2.96 \times 10^{44} \ergps$ 
The luminosity rate of change is $\Delta$L/$\Delta$t $\approx$ 7.8 $\times$
10$^{40}$ erg s$^{-2}$.  Quantifying this rate of luminosity change in terms
of a radiative efficiency, $\eta \gs 4.8 \times 10^{-43}$$\Delta$L/$\Delta$t
(Fabian 1979), we calculate $\eta \gs 0.037$.  This is consistent with the 
efficiency of a Schwarzschild black hole.

A $U$-filter ($3000-3900 \A$) light curve was constructed from the OM data  
during the first $\sim 25$~ks of the observation (lower panel of
Figure~\ref{fig:lc}).  In general, and in particular during the X-ray flare,
the optical light curve shows no evidence of variability.  A constant model 
fitted to the light curve was acceptable ($\chi^{2}_{\nu} = 0.9$).
In addition, two shorter light curves covering $\sim 5$~ks each were
obtained with the $B$ ($3900-4900 \A$) and $UVW2$ ($1800-2250 \A$) filters.  
Neither of these curves showed
any variability ($\chi^{2}_{\nu} < 0.6$ for 4 $dof$ in each filter). 


\subsection{Spectral variablity}
\label{sect:sv}

\begin{figure}
\rotatebox{270}
{
 \scalebox{0.32}{\includegraphics{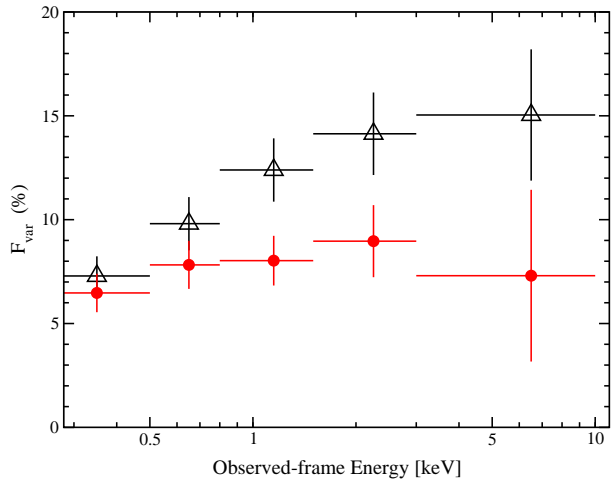}}}
\caption{
The RMS spectra of \nab\ over the entire observation (open, black triangles)
and excluding the flaring event between $\sim 10-15$~ks (red dots).  F$_{var}$
is calculated using light curves in 1000~s bins.
}
\label{fig:fvar}
\end{figure}

The fractional variability amplitude
($F_{var}$; Edelson et al. 2002) was calculated twice (Figure~\ref{fig:fvar}).
Firstly, utilising data over the entire observation, and again while
ignoring the data during the apparent flare ($\sim 10-15$~ks).  As can be seen
from Figure~\ref{fig:fvar} the non-flare rms spectrum is consistent with a
constant ($\chi^{2}_{\nu} = 0.6$).  On the other hand the flare data set 
shows clear spectral variability, with a gradual increase in the amplitude
of the fluctuations with increasing energy.  A constant fit is unacceptable
($\chi^{2}_{\nu} = 6.8$).

Of interest is the notable similarity in the appearance and behaviour of the
rms spectra with the rms spectra of another narrow-line quasar, 
I~Zw~1 (Gallo et al. 2004).
Gallo et al. also observed a flare in I~Zw~1, of similar magnitude and 
duration to that seen in Figure~\ref{fig:lc}.
In addition, it was discovered that the flare in I~Zw~1 was concentrated
in the hard energy band (i.e. 2--10~keV), and that it induced spectral 
variability as suggested by Figure~\ref{fig:fvar}.

\begin{figure}
\rotatebox{270}
{
 \scalebox{0.32}{\includegraphics{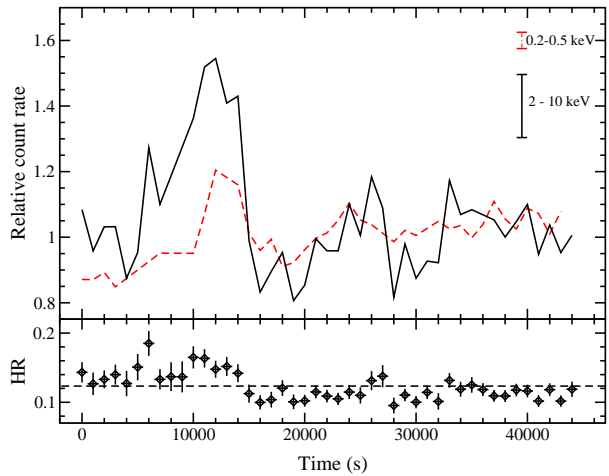}}}
\caption{
Top panel: Normalised light curves in the 0.2--0.5~keV (red, dashed curve)
and 2--10~keV (black, solid curve) bands.  
The light curves were normalised to the
average count rate during the non-flare periods (i.e. excluding the data
between $\sim 10-15$~ks), to demonstrate
the different flare amplitude in each band.  The average size of the 
uncertainties for each curve are shown in the upper right of the panel. 
Bottom panel:  The hardness ratio (HR = hard/soft) of the two light curves
in the top panel as a function of time.  Bin sizes are 1000s.  
A constant fit is drawn in as a dashed line.
}
\label{fig:hr}
\end{figure}

In Figure~\ref{fig:hr} we present normalised light curves in the
0.2--0.5~keV and 2--10~keV bands.  The light curves were normalised to the
average count rate during the non-flare periods (i.e. excluding the data
between $\sim 10-15$~ks).  As in I~Zw~1, the flare appears to be concentrated
in the harder energy band (though much less significant than for I~Zw~1).
The hardness ratio between the two light curves is also presented in 
Figure~\ref{fig:hr}, and shows a difference in the hardness ratio
before and after the flare.

Examining a ``flaring'' spectrum in detail is difficult due
to the small number of good events in the first half of the observation.
The mean spectrum is essentially dominated by the data in the post-flare
regime (after $\sim 20$~ks).


\section{Discussion and Conclusions}
\label{sect:dis}

\subsection{General Findings}

The main results of this analysis are the following:

\begin{itemize}
\item[(1)]
The X-ray continuum of \nab\ was well fitted with a power-law plus blackbody
with $\Gamma \approx 2.45$ and $kT \approx 120$~eV.  Additional absorption
above the Galactic column density was not required. \\

\item[(2)]
A strong, broad feature was detected in the pn spectrum between 5--6~keV.
A considerable amount of the emission, including the peak of the emission, 
is redward of 6.4~keV (rest-frame), and
extends as far down as 5~keV in the rest-frame.
The feature could be fitted as a neutral iron line emitted from a narrow 
annulus of a disc 
around a Schwarzschild black hole.  \\

\item[(3)]
A flare was detected in the light curve in which the broad-band
count rate increased by about 50\% in a few thousand seconds.  The flare
was determined to be concentrated at higher energies (2--10~keV) and
appeared to induce spectral variability. \\ 

\item[(4)]
No optical variability was detected from \nab\, even during the X-ray flare.

\end{itemize}

\subsection{The nature of the high-energy emission line}

Given the energy of the broad feature in \nab\, we considered the possibility
that the broad line could be the composition of several narrower lines.
Means to produce narrow emission lines in the 5--6~keV range have been 
suggested (e.g. Skibo 1997; Dov\v ciak et al. 2004); however the hypothetical
narrow features in \nab\ do not satisfy these models physically (e.g.
lines are too strong, or absence of accompanining features, or 
inconsistent relative abundances).  Moreover, there was no statistical
necessity to split the broad feature into multiple components.

The excess emission at $\sim 5$~keV can be fitted as neutral iron 
emitted from an accretion disc around a Schwarzschild or Kerr black hole. 
The fits cannot distiguish between the different line origins.
In the case of a Schwarzschild black hole the line would originate from a 
thin annulus ($18-24~R_g$) close to the black hole.
The feature could be fitted equally well with a simple Gaussian profile, but
the best-fit energy and width do not correspond with the 
characteristics of any elements in the region.  

\subsection{A comparison to \mcg\ and I~Zw~1}
There are three significant differences in the \xmm\ observation of \nab\
since it was 
last observed with $ASCA$ (Fiore et al. 1998; Leighly 1999a,b; 
Vaughan et al. 1999; Reeves \& Turner 2000).
Firstly, the power-law slope is notably softer in this
observation ($\Gamma \approx 2.45$) compared to the $ASCA$ spectrum
($\Gamma \approx 2.09-2.22$). 
Secondly, there was no detection of a broad emission feature in the 
earlier observations, though at low significance there did appear to be 
spectral hardening with increasing energy (Fiore et al. 1998).
Finally, a (hard) X-ray flare was detected with \xmm.

There was also weak evidence for a temperature change in the blackbody 
component.
The temperature measured by $ASCA$ was between 160--220~eV, whereas
the temperature was much lower in the \xmm\ data ($kT \approx 120$~eV).  
The reliability of the measured change is questionable given that $ASCA$ was 
not sensitive
below $\sim 0.6$~keV, and due to the large range of temperatures 
measured in three different analyses of the same data.
The average 0.6--10~keV \xmm\ flux was about 15\% lower compared to the 
$ASCA$ flux;
however variations on this order are typical on hourly time scales. 

The three new features are reminiscent of the second $ASCA$ observation of
\mcg\ during which part of the time the broad line was found redshifted 
below 6~keV
with no component detected at 6.4~keV, and the continuum appeared unusually
soft (Iwasawa et al. 1999).
Those authors interpreted the strange line emission as arising from an
extraordinarily large gravitational redshift resulting from a flare
occuring within $5~R_{g}$ of the black hole.

There is ample evidence for spectral softening in AGN during periods of increased 
2--10~keV intensity (e.g. Done, Madejski \& Zycki 2000; Vaughan \& Edelson 
2001).  Merloni \& Fabian (2001) have described this in terms of a 
`thundercloud' model where a heating event in a compact region of the corona,
likely due to magnetic reconnection (e.g. Galeev, Rosner \& Vaiana 1979), 
results in an avalanche of
smaller events.  As the hard X-ray emission is reprocessed in the disc
softer spectra are produced and the luminosity of the iron line is enhanced.
The scenario appears applicable here, though it is difficult to examine
in detail given the limited amount of data.  The softer average spectrum
(dominated by the post-flare data), the hard X-ray flare, and the
iron line possibly emitted from an annular region are all consistent with the 
thundercloud model.

In this sense the flare observed in I~Zw~1 (Gallo \et 2004) is different
in that no variability in the power-law slope, or enhancement of the 
iron line emission were detected.  This can still be understood in terms
of the thundercloud model if the I~Zw~1 flare originated closer to the
disc, but farther from the black hole.  Thus, the flare illuminated a smaller
portion of the disc, and spectral changes were smeared out in 
the mean spectrum.  Support in favour of this scenario was the non-detection
of a lag between energy bands in the I~Zw~1 flare.



In this most recent and sensitive X-ray observation of \nab, this
narrow-line quasar has exhibited many differences since last observed
with $ASCA$.  The thundercloud model (Merloni \& Fabian 2001) can be
invoked to explain the hard X-ray flare, soft 2--10~keV power-law,
and broad iron emission.  
Whether we examine narrow-line quasars or NLS1, it is becoming quite clear
that these objects are important to understand the relativistic effects close 
to the black hole.


\section*{Acknowledgements}

Based on observations obtained with \xmm, an ESA science mission with
instruments and contributions directly funded by ESA Member States and
the USA (NASA).  
Many thanks to the referee, Dirk Grupe, for a quick and helpful report.
WNB acknowledges support from NASA grant NAG5-9933.



\bsp
\label{lastpage}

\begin{thebibliography}{}

\bibitem{2} Arnaud K., 1996, in: {\it Astronomical Data Analysis Software and Systems}, Jacoby G.,  Barnes J., eds, ASP Conf. Series Vol. 101, p17 

\bibitem{} Bahcall, N. A., Bahcall J. N., Schmidt M., 1973, ApJ, 183, 777

\bibitem{} Boller Th., Brandt W. N., Fink H., 1996, A\&A, 305, 53

\bibitem{} Boroson T. A., Green R. F., 1992, ApJS, 80, 109 

\bibitem{} Boroson T. A., 2002, ApJ, 565, 78 

\bibitem{} Brandt W. N., Mathur S., Elvis M., 1997, MNRAS, 285, 25 

\bibitem{30h} den Herder J. W., \et 2001, A\&A, 365, L7 

\bibitem{31} Dickey J. M., Lockman F. J., 1990, ARA\&A, 28, 215

\bibitem{32} Dov\v ciak M., Bianchi S., Guainazzi M., Karas V., Matt G.,
2004, MNRAS, 350, 745

\bibitem{} Done C., Madejski G. M., Zycki P. T., 2000, ApJ, 536, 213  

\bibitem{}
Edelson R., Turner T.J., Pounds K., Vaughan S. Markowitz A.,
Marshall H., Dobbie, P., Warwick R., 2002, ApJ, 568, 610

\bibitem{33}
Fabian A. C., 1979, Proc. R. Soc. London, Ser. A, 366, 449

\bibitem{35} Fabian A. C., Rees M. J., Stella L., White N. E., 1989, MNRAS, 238, 729

\bibitem{} Fiore F., Elvis M., McDowell J. C., Siemiginowska A., Wilkes B. J.,
1994, ApJ, 431, 515  

\bibitem{36} Fiore F., \et 1998, MNRAS, 298, 103

\bibitem{}
Galeev A. A., Rosner R., Vaiana G. S., 1979, ApJ, 229, 318

\bibitem{34}
Gallo L. C., Boller Th., Brandt W. N., Fabian A. C., Vaughan S., 2004, A\&A,
417, 29

\bibitem{} Grupe D., 2004, AJ, 127, 1799

\bibitem{40} Guainazzi M., 2003, A\&A, 401, 903

\bibitem{41} Iwasawa K., Fabian A. C., Young A. J., Inoue H., Matsumoto C.,
1999, MNRAS, 306, L19

\bibitem{75} Jansen F. \et 2001, A\&A, 365, L1 

\bibitem{}
Kirsch, M. 2003, {\em XMM-Newton} Calibration Presentations
(CAL-TN-0018-2-1)

\bibitem{} Klimek E. S., Gaskell C. M., Hedrick C. H., 2004, ApJ, 609, 69 

\bibitem{80} Laor A., 1991, ApJ, 376, 90

\bibitem{} Leighly K. 1999a, ApJS, 125, 317

\bibitem{} Leighly K. 1999b, ApJS, 125, 297

\bibitem{}
Mason, K. O., Breeveld, A., Much, R., et al. 2001, A\&A, 365, 36

\bibitem{} Merloni A., Fabian A. C., 2001, MNRAS, 328, 958 

\bibitem{129} Nicastro F., Fiore F., Matt G., 1999, ApJ, 517, 108

\bibitem{} Osterbrock D. E., Pogge R. W., 1985, ApJ, 297, 166

\bibitem{}
Pollock 2003, {\em XMM-Newton} Calibration Presentations
(CAL-TN-0030-2-1)

\bibitem{}
Reeves J. N., Turner M. J. L., 2000, MNRAS, 316, 234

\bibitem{129a}
Skibo J. G., 1997, ApJ, 478, 522

\bibitem{165} Str\"{u}der L. \et 2001, A\&A, 365, L18 

\bibitem{} Sulentic J. W., Zwitter T., Marziani P., Dultzin-Hacyan, D.,
2000, ApJ, 536, 5

\bibitem{166}
Turner M. J. L., Abbey A., Arnaud M.,
 Balasini M. Barbera M., Belsole, E.,
 Bennie P. J., Bernard J. P., Bignami G. F.,
 Boer M. et al., 2001, A\&A, 365, 27

\bibitem{177a} Turner T. J. \et 2002, ApJ, 574, L123

\bibitem{177} Turner T. J., Kraemer S. B., Reeves J. N., 2004, ApJ, 603, 62

\bibitem{}
Vaughan S., Reeves J., Warwick R., Edelson R., 1999, MNRAS, 309, 113

\bibitem{}
Vaughan S., Edelson R., 2001, ApJ, 548, 694  

\bibitem{167a}
Yaqoob T., George I. M., Kallman T. R., Padmanabhan U., Weaver K. A.,
Turner T. J., 2003, ApJ, 596, 85


\end{thebibliography}
\end{document}